# Development of Display Ads Retrieval System to Match Publisher's Contents

[1]Izuddin Zainalabidin, [2]Izyan Izzati A Halim, [3]Faizal Ahmad Fadzil,
[1,3]Department of Computer and Information Sciences, Universiti Teknologi PETRONAS,
31750 Bandar Seri Iskandar, Perak, MALAYSIA
[1,3]E-mail: {izuddin_z, faizal_ahmadfadzil}@petronas.com.my

*Abstract*
*The technological transformation and automation of digital content delivery has revolutionized the media industry. Advertising landscape is gradually shifting its traditional media forms to the emergent of Internet advertising. In this paper, the types of internet advertising to be discussed on are contextual and sponsored search ads. These types of advertising have the central challenge of finding the "best match" between a given context and a suitable advertisement, through a principled method. Furthermore, there are four main players that exist in the Internet advertising ecosystem -- users, advertisers, ad exchange and publishers. Hence, to find ways to counter the central challenge, the paper addresses two objectives: how to successfully make the best contextual ads selections to match to a web page content to ensure that there is a valuable connection between the web page and the contextual ads. All methods, discussions, conclusion and future recommendations are presented as per sections. Hence, in order to prove the working mechanism of matching contextual ads and web pages, web pages together with the ads matching system are developed as a prototype.*

## 1. Introduction

Advertising is a marketing form of communication for businesses to promote their products and services to their audience. The technological transformation and automation of digital content delivery has revolutionized the media industry, causing the Internet turning into an advertising channel rapidly. Hua, Mei, and Li (2008) highlighted that the Internet advertising has embarked on a dramatic evolution, which will be rapid, fundamental, and permanent. According to Interactive Advertising Bureau's (IAB) recent report, in the first half of 2012, Internet advertising revenues climbed to an all-time high of $17 billion, representing a 14 percent increase year-over-year. This represents a 14% (or $2.1 billion) increase from 2011's $14.9 billion. Kumar (2012) studied that the Internet spending in Malaysia, one of Asia Pacific countries, and concluded that the trend would continue to grow strongly for the rest of the year, and as per estimates by IPG Mediabrands (Asia world market) it has been growing at similar rates for the last two years.

Contextual Advertising, instead of displaying the same ads to everyone, different ads are shown with regards to the geography, language, device and other characteristics of visitors, to maximize the utilization of advertising opportunities. There are three different types of contextual ads: separate ads that appear in specific areas on a page, inline or in-text contextual ads, or pop-up ads. Some recognizable contextual advertisers are Google's Adsense, Overture's Content Match and AdBrite. Small to medium business owners can save expenses by using a contextual advertising program because the targets are so specific. Certain sites, like news and publisher sites benefit from contextual advertising because the content on these sites is so specific that the returned ads will be targeted to the search engine user. The common problems with online advertising are:

- Modeling Internet ads that likely match user's interest, whereby the challenge comes in when advertisers need to know user's preferences at the first place, in order to display the most relevant and appropriate ads that match to user's specific interest.
- Establishing Internet ads that are more likely to be clicked on, where publishers need to find a technique in establishing the suitable ads closer to the content served on the Web, so that the ads match user's preferences and eventually will increase the possibility of the ads being clicked by the users when they are online.

- Minimizing advertising costs for advertisers with low budget in marketing their products, where advertisers need to successfully display their ads that match to user's preferences, but at the same time, generating as much revenue as possible.

The objectives of this paper are to successfully make the best contextual ads selections that match to the web page contents through the concept of Computational advertising.
- To ensure there is a valuable connection between the Web pages and the contextual ads.

## 2. Related Work

Advertising is a form of communication used to encourage or persuade an audience (viewers, readers or listeners) to continue or take some action. Traditionally, advertising has been defined as a form of controlled communication that attempts to persuade consumers, using strategies and appeals, to buy or use a particular product or service (Defleur & Dennis, 1996). The advertising industry has a long history of creativity and innovation. Nowadays, Internet and mobile phone technology is in the process of transforming the way societies not only communicate with each other, but introducing many new platforms of Internet advertising. According to Zinkhan and Watson, (1996), new ideas and technologies spread rapidly in Western- style democracies, in part due to the advanced nature of the communication industry, including advertising. In the 1990's, consumers perceived online advertising with banner, pop-up, and e-mail advertisements.

Today, online advertising includes not only evolved banner, pop-up, and e-mail ads but also search marketing, sponsored search, pay-per-click, pay-per-action, rich media, Contextual Advertising, geo-targeting, behavioral targeting, social marketing, video advertising, and user-generated online video. Advertising also is appearing in online games, in-line text, social media, blogs, and mobile formats (Boone, Secci, and Gallant, n.d.). In one of Asia Pacific countries, Malaysia, there have been as many as seven new online advertising network companies entering the local scene. According to Tan (2009), this proves that online advertisers realize the importance of ad networks in their digital media plan mix and are allocating more budget to ad networks as they provide higher audience reach in the relevant environment and give better ROI (return on investment) with media or creative optimization technology.

Hence, digital media in Malaysia are on the edge for "explosive growth" and are finally gaining credibility as mainstream – as opposed to niche (classic advertising methods). Furthermore, according to MCMC, the digital world is changing not only consumers' habits in Malaysia, but also encouraging advertising opportunities, especially online advertising. Therefore nowadays, advertising industry has evolved from classic/traditional advertising channels (radio, television, print, billboards, outdoor advertisements), to Internet advertising due to the technology advancements, and also due to the consumers' habit. Moreover, the developments of computer mediated communication technology, brings up the opportunity for marketer to get direct feedback from consumers and Website visitors, along with the ability to customize individual messages (Haque, et al. 2007). It is believed that soon in the future, Internet advertising will change the current trends and starts to dominate the existing media as the preferred medium for placing ads.

In this study, since the types of advertising that are being focused on (sponsored search and contextual Ads) require computation and principled way of finding the best match between a user in a given context and available ads, they are also referred to as computational advertising. Computational advertising is an emerging scientific discipline, at the intersection of large scale search and text analysis, information retrieval, statistical modeling, machine learning, optimization, and microeconomics. (Broder, 2008).

According to Dave (2011), computational advertising is about using various computational methodologies to do contextually targeted advertising. As discussed, computational advertising mainly addressed: retrieving a set of ads that best matches the context and then ranking these ads. Dave (2011) proposes that sponsored search refers to the placement of ads on search results page. Here, the context is the query issued by the user and the problem is to retrieve top relevant ads that semantically match the query. Contextual Advertising deals with the placement of ads on third-party Web pages. It is similar to sponsored search, with the ads being matched to the complete web page text as opposed to a query.

There are two key technological innovations that trigger the sponsored search process, both of which depend on advanced economic and mathematical methods, and which ultimately help determine the nature of the market structure. They include: 1) Search-ad platform using "keyword bidding system" 2) Maximizing revenue from selling slots (David S. Evans, 2008). In contextual advertising, instead of displaying the same ads to everyone, different ads are shown with regard to the geography, language, device and other characteristics of visitors, to maximize the utilization of advertising opportunities (Yuan, et al. 2012). Contextual advertising, such as Google AdSense, is a type of text-based search ad which often presents three to five different ads in a frame, each including title, description, and display URL (Yung-Ming Li and Jhih-Hua Jhang-Li, 2008). According to Broder, et al. (n.d.), In Contextual Advertising usually there is a commercial intermediary, called an ad-network, in charge of optimizing the ad selection with the twin goal of increasing revenue (shared between publisher and ad-network) and improving user experience.

There are four major players (Yuan et al, 2012) in the Internet advertising ecosystem, which are: users, advertisers, publishers and ad exchange. Each and every player in the ecosystem has their own utilities that need to be considered.

User is a person who browses the Web, consumes media content and performs searches using search engines. It also can be said that user is a person that visits the Web pages of the publisher and interacts with the ads (Broder, et. al, n.d.). Hence, the utility of a user is, having to be provided with suitable ads that best match their preferences and interest, while he/she is online. While browsing on the Internet, users seem to acknowledge that the presence of ads is what allows them to view much of the content on the Web for free and have accepted Web advertising as a way of life on the Internet (McDonald and Crano, 2009). Nevertheless, Enquiro (2008) verified that users show a reluctance and unwillingness to click on Web advertisements and make an active effort to avoid doing so, viewing advertising as a visual obstruction, impeding the content of the page. To further exemplify this point, Jansen and Resnick (2006) demonstrate that users performing e-commerce searches were more likely to click on a non-sponsored link for a Website than the same link contained in an advert.

Advertiser provides the supply of ads. Usually the activities of the advertisers are organized around campaigns which are defined by a set of ads with a particular temporal and thematic goal (Broder, et al. n.d.). In order to generate demands for its products or service form its customers, advertiser will venture into commercializing ads that will attract more of its customers to buy the products and services offered. Thus, the main utility of an advertiser in Internet advertising ecosystem would be maximizing the click-through-rate (CTR) by viewing each impression or click as an asset with future returns. In most networks, the amount paid by the advertiser for each sponsored search click is determined by an auction process where the advertisers place bids on a search phrase, and their position in the tower of ads displayed in conjunction with the result is determined by their bid (Broder, et al. n.d.). Usually, advertiser participates in keyword auctions, where keywords are selected before the auction starts. The keywords selection must be done carefully, in order to keep up with the budget, and at the same time, profits are still at the maximum point. As the keyword statistics change from time to time, making keyword selection can be overwhelming. Apart from that, Naldi, et al. (2010), explore the application and implications of the use of the generalized second-price (GSP) auction mechanism for sponsored link slot assignment and pricing in sponsored advertisement keyword auctions. The GSP's auction mechanism allocates advertising slots in the descending order of advertisers' bid prices. Those advertisers who bid the highest prices are given the most attractive sponsored advertising slots at the top of the Web Page, where the click-through rates have been shown to be the highest (Brooks, 2004).

Publisher is the owner of the Web pages on which the advertising is displayed. The publisher typically aims to maximize advertising revenue (Return of Investment (ROI)) while providing a good user experience (Broder, et. al., n.d.). In order to make the most money, the publisher will make the best use of its available ad inventory. Therefore a challenge for publishers is to select the optimal contract or estimate the optimal price (Yuan, et al. 2012). Publisher is on maximizing the revenue from the over –the-counter (OTC) contracts. This is due to the strategy here is more perceptible as when the publisher received requests directly from the advertisers, the publisher will have more control on it. Work by Roels and Fridgeirsdottir, (2009) and Feige, et al. (2008) incorporated contract guarantees, where an attempt was made to maximize a publisher's revenue in display ads through dynamic optimization. Apart from that, publisher also may schedule and improve its content quality in order to maximize their revenue and profits return. Nakamura and Abe (2005) developed an LP-based algorithm to schedule banner ads, where they presented three features that each ad was associated with; 1) time of day that the ads were preferred to be viewed (e.g. the afternoon), 2) page category (e.g. sports) and 3) the number of impressions. Optimal ad time and location that maximizes overall revenue are successfully determined by using these three features, rather than relying only on the CTR of an individual ad. It is proved that their strategy showed an enhancement over greedy and random methods.

Increasingly, display advertisements (ads) on the Internet are sold via marketplaces that bring publishers and advertisers together in real time when an opportunity arises to present an ad to a viewer (Muthukrishnan, 2009). These marketplaces are Ad Exchange (ADX). According to Yuan, et al. (2012), the ad exchange (ADX) is considered a uniform marketplace for publishers to sell ad inventories, and for advertisers to buy impressions and clicks. In simple words, ADX serves as the interface between publishers and advertisers in the Internet advertising ecosystem. Therefore, to ensure the valuable connection between publishers and advertisers, an ADX needs to fulfill the twin goal of increasing revenue (shared between publisher and ad-network), and improving user experience. Nowadays the generalized second price auction (GSP) is the most adopted model in ADX (Yuan, et al. 2012). To add to that, Edelman, et al. (2007)

demonstrates that GSP is when the advertiser pays the next highest bid instead of their own bid price. In addition, due to the probability that an ad will be clicked (CTR) will determine the value of the ADX revenue, ADX needs to consider the relevancy. If the relevancy is less in value, users will retreat (due to less satisfaction), and refuses to come back. In a follow-up work (Lacerda, et al. 2006) the authors propose a method to learn impact of individual features using genetic programming to produce a matching function.

Furthermore, Neto, et al., (2005) proposed to generate an augmented representation of the target page by means of a Bayesian model built over several additional Web pages. Besides, Radlinski et al. (2008) proposed an online query expansion algorithm of two stages: the offline processing module pre-computes query expansions for a large number of queries, and then builds an inverted index from the expanded query features. Later, Broder et al. (2008b) proposed a method for both augmenting queries and ads. Three distinct spaces of different features are used to represent queries: unigrams, classes and phrases extracted using a proprietary variant of Altavista's Prisma refinement tool.

Targeted or contextual ad insertion plays an important role in optimizing the financial return of this model (Yang, et al. 2010). Contextual advertising is a type of Internet advertising used for content- based Web sites that make targeted ads appear on a Web page based on the page's actual content. Typically, a code snippet is placed on the page, and the code "figures out" what the page is about and serves appropriate advertisements from a large database of advertisers. This type of advertising works really well for Website owners because the link titles are relevant to the content and they do not get ignored like most advertising.

In addition, Lang (1995) describes a system which separates advertisements and publishers' Web sites by introducing an advertisement agent. The agent sits between advertisers and the user's browser and merges banner advertisement directly into the currently viewed page, independent of the page itself. According to Frank, et. al. (1999), KEA keyphrase extraction algorithm, extracts keywords using a simple machine learning mechanism. It shows a simple procedure for key phrase extraction based on the naive Bayes learning scheme performs comparably to the state of the art.

On the other hand, Yih, et al. (2006) made a great improvement over KEA. The general architecture of the keyword extraction system consists of the follow four stages: *Preprocessor, Candidate Selector, Classifier, and Postprocessor.* Apart from that, Ribeiro-Neto et al., (2005) describes an Impedance Coupling technique for content-targeted advertising which expands the text of the Web page to reduce vocabulary impedance with regard to an advertisement, can yield extra gains in average precision of 50%. For this, they proposed to expand the triggering pages with new terms. Most importantly, their work focused not on finding keywords on Web pages, but on directly matching advertisements to Web pages.

Other works have addressed the ads targeting issue includes a proposed system called ADWIZ that is able to adapt online advertisement to a user's short-term interests in a non-intrusive way, suggested by Langheinrich et al., (1999). Contrary to the work done by Ribeiro-Neto et al., ADWIZ does not directly use the content of the page viewed by the user. It relies on search keywords supplied by the user to search engines and on the URL of the page requested by the user. Meanwhile, Lacerda et al. (2006) proposed to use machine learning to find good ranking functions for contextual advertising. They use the same dataset described in the paper by Ribeiro-Neto et al. (2005). Lacerda et al. (2006) find that the ranking functions selected in this way are considerably more accurate than the baseline proposed in Ribeiro-Neto et al. (2005). Additionally, according to Anagnostopoulos et al. (2006), for static pages that are displayed repeatedly, the matching of ads can be based on prior analysis of their entire content; however, often ads need to be matched to new or dynamically created pages that cannot be processed ahead of time. Thus their work focused on the contributions of the different fragments of the pages.

Murdock et al. (2007) consider machine translation to overcome the vocabulary mismatch between target pages and ads. In more detail, the machine translation features they use correspond to the average translation probability of all words in the target page translated to the keywords or to the description of the ad, and the proportion of translations of the ad terms, or the ad keywords, that appear on the target page. On the other hand, Chakrabarti et al. (2008) proposed a new class of models to combine relevance with click feedback for a contextual advertising system. Their model is based on a logistic regression and allows for a large number of granular features. Broder et al. (2007) notice that the standard string matching approach can be improved by adopting a matching model which additionally takes into account topical proximity. In their model the target page and the ad are classified with respect to taxonomy of topics.

Besides that, Fan and Chang (2009) put forward their ideas with regards to matching contextual ads and Web page through a novel framework for associating ads with blog pages based on sentiment analysis. More to add, Hatch et al. (2010) introduced "clickable terms" approach to contextual advertising. This approach involves matching a Web site directly with a set of ad-side terms, independent of the page content. They use log-likelihood ratios (LLRs) to measure the relative click-ability of a given ad-side term on a given site. Other than that, Wu et al. (2011) suggested a new approach by incorporating the Wikipedia concept and category information into the traditional keyword matching to enrich the content representation of pages and ads. They described how to map each ad (or page) into a keyword vector, a concept vector and a category vector, as well as how to combine the three feature vectors together for making the top-N ads selection.

Recent views on this matching system had been done by Joshi et al. (2013) where they make contextual targeting more relevant with Extraction of relevant entities from the Web page. They extract the entities from Web page, which is of interest to the consumer. Then they target the interest of Internet user and put up the ads according to their interest. The system is designed in such a way that it can extract entities (e.g. Name, Place, Title, Location, Date) from Web page and ad publisher put up a advertise on that page which include those entities which are extracted from page.

## 3. Development of Display Ads Retrieval System

In this study, the development tools used to develop the system are OpenX. Apache HTTP Server, Wordpress, and Adobe Photoshop CS5.

### 3.1 OpenX

OpenX is an ad server that can be used to manage and optimize the advertising space on one or more websites. In this project, this tool is used to manage the ad server and measure the statistics of ad banners that are placed on the Web pages. OpenX acts as the connector between publisher and advertisers, giving out the same roles as ad exchange.

### 3.2 Apache HTTP Server

Apache HTTP Server is an open-source Web server platform. In this project, this tool is utilized to deliverer Web pages through the Internet. When the Apache program receives the request for a file, it looks for the file on its disk, and when found, sends that file to the requester in a stream of data named HyperText Transfer Protocol, which is then decoded by the browser program and rendered as a Web page on the requester's screen.

### 3.3 Wordpress

In this project, Wordpress is used in the development of different Web pages. It is a semantic personal publishing platform with a focus on aesthetics, Web standards, and usability. WordPress is completely customizable and can be used for almost anything.

### 3.4 Adobe Photoshop CS5

Adobe Photoshop CS5 is a graphic-editing program developed and published by Adobe Systems. In this project, this tool is used to design the content and user-interface of Web pages. Besides it is also used to design ad banners that are stored in the ad server in the system.

Figure 1 shows that the system has three main elements that will serve the system purpose which are ad server, browser and publisher.

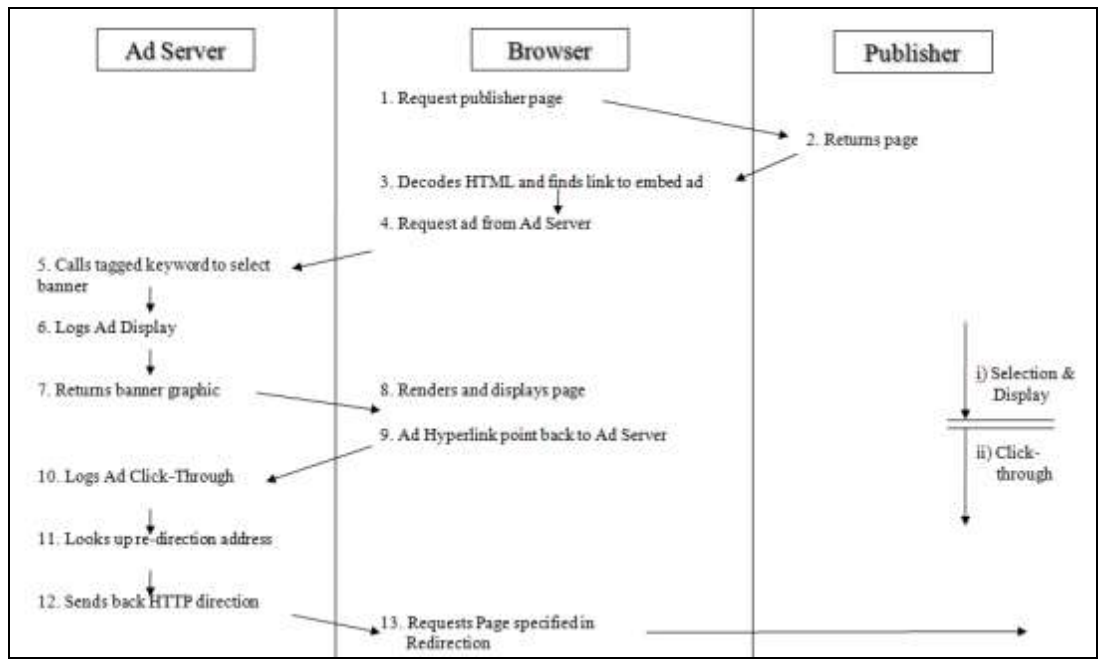

Figure 1: System Flow Diagram

Ad server stores all of the ads and delivers them to the requests made by the Web pages through Web browser. Meanwhile Web browser displays the delivered ads form the ad server to the user and publisher returns the requested page from the Web server. More to add, Web server holds all the database of the Web pages and Web pages contents. Also, Web server stores the database of the ad server. Hence the ad server is the subset of the whole Web server. In system modeling phase, the author had model some of the system characteristics and features in order to recognize the behavior of the system as it evolves over the development period. This model will form a set of assumptions concerning the operation of the system. This model also used to study systems in the design stage before the system is being built. Hence, Table 1 states the system and its components:

Table 1: Ads matching System and Its Components

| Entities | Attributes | Activities | Events | State Variables |
| --- | --- | --- | --- | --- |
| Ads; Web pages | Ad keywords and tags; Web page contents | Matching ads with web page contents | Ad requests; ad delivery; ad display | Type of web page contents; Ads tag; Ads schedule |

The modeling phase of the system ranged in three phases based on the behaviors and features of the Web pages and the ads delivery:

- Static Web page contents and static ads delivery method;
- Static Web page contents and dynamic ads delivery method;
- Dynamic Web page contents and dynamic ads delivery method.

Thus, it is supposed that this third phase of the modeling will simulate 99% of the real-life situations of the system. In simulating the real environment of ads matching system and getting the best match between contextual ads and the Web page, the author proposed to have an ads matching system that have an agent between publisher and advertiser in the Internet advertising ecosystem. The agent in between will act as a platform to serve publisher and advertiser in advertising their ads and to get more targeted customers that will likely click on their advertisements. Hence to set up this agent in the middle, the author used an ad serving management ad server which is OpenX Ad Server.

## 4. Results and Analysis

A survey was conducted mainly aimed to identify what are the Internet users' perspectives and views on Internet advertising. The survey is conducted online, targeting responses from 65 Malaysian Internet users, specifically. There are 22 close-ended questions being asked in the survey. There are two sections of questionnaires: 1) Demographic Data 2) About Internet Advertising. Based on this online survey, Figure 2 shows that more than 50% of Internet users click on the Internet ads because they found that those ads matched to their interests and preferences:

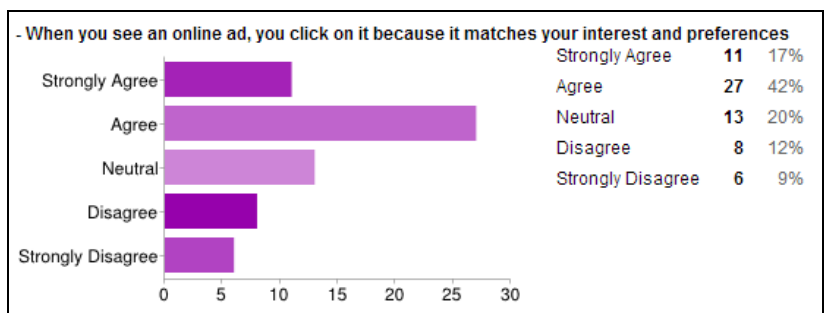
Figure 2: Survey analysis

The OpenX Ad Server is integrated with three Web pages. All three web pages are developed by using Wordpress and its customized templates. Three Websites under the theme 'Photography' are created to demonstrate ads-matching varieties. Scoping down the 'Photography' theme, the first Website (named The Picstop) is mainly about giving out reviews to latest gadgets for photographers like cameras, lenses and camera accessories. The second Website (named The Shutter Up Photography) is a photography-service Website, whereby the owner of the Website is providing portrait photography service to its clients. The third Website (named Bridalsnaps) also provides photography service, but they are specializing in bridal photography, and taking photos for wedding receptions. Besides, Bridalsnaps also provides wedding photography tips.

The owner of these Websites will then act as publishers. Later, the ads (contextual ads) paid by advertisers will be displayed on their Websites will mainly related to their Website contents, satisfying the relevancy of the displayed ads. Hence the final step is to link the right campaigns to the right zones, effectively determining which ads will appear where. This can be combined with various forms of targeting, matching the display of ads to specific situations. OpenX Ad Server can handle targeting based on date, day and time, geographic area (country or city, and more), Web browser used by the visitor, and language set by the visitor, to name just a few.

To prove the concept, advertisers, campaigns and ads in the OpenX Ad Server are created according to the campaign demand. In OpenX Ad Server, an advertiser has a name, a contact and their e-mail address. Any advertiser can have one or more campaigns. A campaign is a collection of ads that have several properties in common. In this project simulation, author used contract campaign type. By default, campaigns have no start and end date. They will start delivering (when linked to zones) straight away after they have been created, and they will not expire. When a campaign has a start date in the future, OpenX Ad Server will automatically activate that campaign at midnight of that date. When a campaign has an end date, the campaign will continue to run until midnight of the date. So both the start date and the end date are inclusive.

In the OpenX Ad Server, the Websites and one or more zones for each Website are defined. A zone represents a space on the Web pages where ads are supposed to be displayed. For every zone there is a little snippet of HTML code, which must be placed in the site, at the exact spot where the zone should go. This forms the integration between the site and the OpenX Ad Server. It is a one time job that will take little time for an experienced Webmaster or developer. The word 'zone' is a term in OpenX, other systems refer to it as 'location', 'spot', 'placement' or 'position'. A zone in OpenX Ad Server has a few properties: name, description and size (width and height). And after creation, every zone gets a unique ID.

Local banners are used for the purpose of experiment. The banners are stored in the ad server's database as the author's perception is to create a middle party between publishers and advertisers. Once advertisers, campaigns and zones are created, it is time to link zones with Websites. In addition to linking campaigns to zones, the author can also link an individual ad to a zone. The ability to link individual ads rather than whole campaigns provides extra flexibility, but it comes at the price of requiring more work to set everything up. OpenX Ad Server generates snippets of html or JavaScript code, called 'invocation code' in OpenX. These snippets of code, also referred to as 'tags', are pasted into the template, which enables the pages that use these templates to show zones and the campaigns and ads linked to those zones. Hence, after that, the author would need to paste this code on the section where advertisers want the advertisements to appear (with the agreement of publishers). The banner code is pasted on the sidebar widget (titled: Sponsors) in the Wordpress admin page. After the banner code has been saved in Wordpress, a user would wait for around 10 minutes, if he or she does not see the advertisement served immediately. Then, refresh the website again. Finally, the advertisement should appear, assigned from OpenX Ad Server.

As illustrated in Figure 3, Ads displayed on the web page are pulled from OpenX Ad Server that match the dynamic Website contents as per assigned. When a user clicks on these ads, they will be redirected to advertisers' landing page automatically.

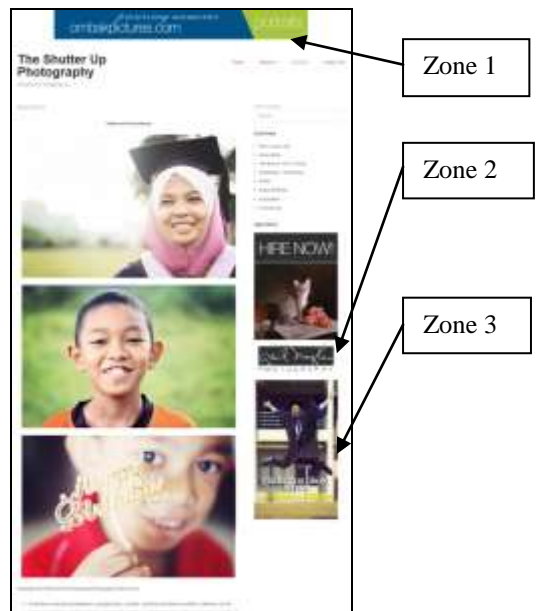

Figure 3: Prototype of one of the websites created

Notice that all ads that appeared on the Website are related to the Website content. As for instance, the first zone (Leaderboard) shows ad from advertiser ombakpictures.com. The ad highlights the portrait photography service it offers at ombackpictures.com. This is related to the Website content since The Shutter Up Photography is also doing portraitures photography service. The same thing applies for ads in zone 2 and zone 3. Both ads are offering photography service. When a user visits The Shutter Up Photography Website, he/she also being served with ads that are related to the Website. Thus the likeability of the user to click on the served ads will be high, as it matches their preferences.

In addition, when ads are matched to user's preferences, the user will perceive ads with high value and not easily being ignored. Furthermore, the same mechanism is used for the other two websites – The Picstop and Bridalsnaps. All ads that are displayed mainly related to the website content. For The Picstop Website, all zones advertised camera products and accessories. As for Bridalsnaps, all zones are set to advertise ads that are mainly related to wedding, such as wedding planning, wedding photography and bridal products and accessories. Many sites also have different templates for different sections, like the home page, the news pages and the forum pages. Whatever the case, the zones are part of the template, and as a result will automatically appear on all the site's pages using that template.

Meanwhile, OpenX Ad Server is capable of targeting ads to specific sections of a Website, using a feature called the 'source parameter'. Source is a way of 'labeling' a zone. The word entered in the source field is used to determine which ads will be displayed. Targeting is a great way to increase ad space revenue. Giving advertisers the opportunity to target specific sections of the Websites could help you to make better use of your ad space and at the same time help you set higher advertising rates. There are also advanced targeting options, using known information about members or customers (like their gender, education or age range), or from known details about the site's content and structure (for instance: display the ad only in the 'news' section or anywhere but the 'sports' pages).

OpenX Ad Server continuously records every single ad impression and ad click. The data it collects is kept in back end of the database, and is not visible through the OpenX Ad Server user interface. The OpenX developers refer to this data as 'raw data', because it hasn't been summarized into the on-screen statistics. In most cases, there is more than one zone on any single Web site, so viewing a Web page will cause multiple ad impressions being logged. The logging of raw data is called 'bucket logging'. Raw data is being recorded for every combination of an ad and a zone, and the software keeps a running total. The first impression of an ad in a zone creates a record with an impression count of 1. The second impression increases that count to 2, and so on.

Thus in a nutshell, OpenX Ad Server is written in the PHP programming language and stores data using the MySQL database, both of which are also open source software. Author has Apache HTTP Server (that comes with MYSQL database) installed in building the Websites and integrating them with OpenX Ad Server. Except for time and hardware, there are no additional costs, and no license fees. The system has been designed to enable installation on a cluster of Web servers, allowing it to grow alongside the growth of the sites.

## 5. Conclusion

In conclusion, this paper focuses on how to successfully make the best contextual ads selections that match to the Web Page contents. Secondarily, it focuses to ensure there is a valuable connection between the Web pages and the contextual ads during the selection process. It is believed that Internet advertising has emerged as the potential channel for e-commerce companies to reach their target customers in marketing and promoting their products. Furthermore, addressing the issues raised in this project is very essentials to make the best Contextual Ads selections that match to the Web Page contents of the business so that ads are more likely to be clicked on by the users. More to the point, computational advertising theory could be applied in creating more effective Internet advertisements in online advertising that are more targeted and best-matched with the Website contents.

Thus, this research is another contribution to the Advertising Industry itself, to be exploitable and acceptable by the local community and contributes to bringing forward the potential of Internet advertising in Malaysia.

In recommendations, future work in this area needs to be done in order to refine the potential of establishing computational advertising, as it is believed that this area of business is likely to propel in the local market in the near future. Besides, future analysis and research on balancing goals for online advertising market players need to be in continuation to enhance the credibility and explore the possibilities of Internet advertising, to be implemented in the local scene. It would be very great if future works will explore more features on making content matching more personalized compared to what is done in this project.

Also, it is suggested that this area of research will take a migration towards behavioral targeted advertising as it will likely boost up online advertising performance. Behavioral Targeting refers to a range of technologies and techniques used by online Website publishers and advertisers which allows them to increase the effectiveness of their campaigns by capturing data generated by Website and landing page visitors. This area of research needs more attention on consumers'' behavior analysis, Human-Computer Interaction (HCI) and consumers' privacy.